\documentclass[journal=jacsat,manuscript=article]{achemso}

\usepackage[version=3]{mhchem} 
\usepackage[utf8]{inputenc}
\usepackage{graphicx} 
\usepackage{amsmath,amssymb,subfigure}
\usepackage{epsfig}
\usepackage{float}
\usepackage{xcolor}
\usepackage{physics}
\usepackage{mathtools}
\usepackage{graphicx}
\usepackage{multirow}
\graphicspath{ {./Figures/} }
\usepackage{multicol}
\usepackage{color,soul}

\author{Abdul Kalam}
\email{ak6994231@gmail.com}
\author{Ashok Kumar}
\affiliation[First University]{Central University of Punjab, Bathinda, India}
\affiliation[Second University]{Department of Physical Sciences, Indian Institute of Science Education and Research Kolkata, India}
\author{Prasanta K. Panigrahi}
\email{pprasanta@iiserkol.ac.in}
\affiliation{Department of Physical Sciences, Indian Institute of Science Education and Research Kolkata, India}

\title[An \textsf{achemso} demo]
  {Study of the Bound State, Electron-Detachment Energy and Reactivity of Hydride Ion using Variational Quantum Eigensolver }

\abbreviations{IR,NMR,UV}

\begin{document}

\begin{tocentry}

Some journals require a graphical entry for the Table of Contents.
This should be laid out ``print ready'' so that the sizing of the
text is correct.

Inside the \texttt{tocentry} environment, the font used is Helvetica
8\,pt, as required by \emph{Journal of the American Chemical
Society}.

The surrounding frame is 9\,cm by 3.5\,cm, which is the maximum
permitted for  \emph{Journal of the American Chemical Society}
graphical table of content entries. The box will not resize if the
content is too big: instead it will overflow the edge of the box.

This box and the associated title will always be printed on a
separate page at the end of the document.

\end{tocentry}

\begin{abstract}
The accurate prediction and understanding of molecular energy and chemical reactivity are fundamental pursuits in the field of molecular quantum chemistry. With the limitations of the current noisy intermediate-scale quantum computer (NISQ) era, the Variational Quantum Eigensolver (VQE) algorithm offers a promising approach to efficiently estimate the stable-energy state of a given molecule.
This study is focused on predicting the ground state and single-electron detachment energy of the hydride ion because of its high electron-electron correlation and numerous applications in astrophysics and quantum chemistry. The hydride ion is the first three-body quantum problem for which the ground-state energy has been calculated theoretically using the "Chandrasekhar Wavefunction," which takes high electron-electron correlation into account. For the calculation of hydride ion, we used the VQE algorithm with two types of quantum computational ansatz, (i) chemistry-inspired ansatz  based on unitary-CC (UCC) and (ii) hardware-efficient based 
 ansatz (HEA).
 To check the versatility of the algorithm, we analyzed two proton transfer reactions involving the hydride ion and finds the energy to be exothermic with a much-improved result compared to previous studies. 
Overall, the VQE algorithm with a quantum computational approach is found to be reliable in calculating stable state energy, single-electron detachment energy, and reaction energy for proton transfer reactions involving a highly correlated molecule with a low relative percentage error.

\end{abstract}

\section{Introduction}


Quantum computation uses principles of quantum mechanics to perform computations that are beyond the capabilities of classical computers \cite{feynman2018simulating,lloyd1996universal,deutsch2020harnessing}. Quantum computers
(QCs) evaluate quantum circuits or programs in a manner similar to a classical computer, but quantum algorithm’s ability to leverage superposition, 
and entanglement is projected to give QCs a significant advantage in cryptography \cite{plesa2018new}, quantum chemistry \cite{mcardle2020quantum,cade2020strategies,azad2022quantum}, optimization \cite{orus2019quantum,li2020quantum,ajagekar2019quantum}, and machine learning  \cite{benedetti2019generative,huang2021power}. In particular, the ground state of a many-body interacting electronic Hamiltonian and predicting the chemical reactivity of molecules is one of the important application areas of quantum chemistry where quantum computation may play a very important role \cite{cerezo2021variational}.

In order to calculate molecular energies, the Hartree-Fock (HF) method which uses a mean-field approach to approximate ground state energy efficiently (resource requirement scaling in $O(N^{4})$) has been commonly used, but without including electronic correlation effects. For higher accuracy, coupled-cluster (CC) methods are often used \cite{bartlett2007coupled}, which converge systematically towards the exact full configuration-interaction (FCI) \cite{sherrill1999configuration} result. The CCSD (Coupled cluster with singles and doubles excitation) method truncates the CC expansion at the doubles term for a balance between efficiency and accuracy \cite{helgaker2013molecular,purvis1982full,dutta2018lower}. The CCSD(T) method, which adds a perturbative treatment of triple excitations (scales in $O(N^{7})$)\cite{bartlett2007coupled}, is considered the "gold standard" method in quantum chemistry for providing high accuracy for most practical purposes \cite{raghavachari1989fifth}. However, finding exact solutions on a classical computer would require exponential resources.

The Phase Estimation Algorithm (PEA), initially proposed for the electronic structure of molecules on a quantum computer  \cite{saue2020dirac,abrams1999quantum}, requires long coherence times and complicated quantum gates which are beyond the capabilities of current quantum devices. To address this issue and make use of the noisy intermediate-scale quantum (NISQ) devices, the variational quantum eigensolver (VQE) algorithm was developed \cite{peruzzo2014variational,mcclean2016theory}. VQE uses fewer qubits and simpler quantum gates, making it suitable for current and upcoming quantum processors. The form of the ansatz is a crucial ingredient of VQE to determine its success on NISQ
devices \cite{google2020hartree,grimsley2019adaptive,yordanov2021qubit}.

The electrons with opposite spins present in negative hydrogen ion (hydride ion) make it an ideal system to study the electron correlation.  \cite{hylleraas1929neue,williamson1942negative,chandrasekhar1944some}. Hydride ion forms a bound structure with low energy electrons in the presence of hydrogen  \cite{rau1996negative}. 
Despite being earlier believed to be unstable, the hydride ion is now known to be present in the solar spectrum, and its absorption is believed to be the primary cause of opacity in the Sun's atmosphere \cite{valley1965handbook}.
Hydride ion is a highly reactive and energetic chemical species with unique properties that are crucial for electric energy storage, utilization, and chemical processing including fuel production \cite{mohtadi2016renaissance}.  In energy-related studies, hydrides have shown remarkable characteristics and capabilities, such as providing both electrons and protons \cite{shima2017dinitrogen} for converting dinitrogen to ammonia through various pathways, such as thermal, biological, organometallic, and electro/photochemical reactions \cite{wang2020power}.

Furthermore, the single-electron detachment energy (also called electron affinity of hydrogen) of the hydride ion has been used as a descriptor in different fields \cite{millar2017negative}, e.g., modeling the chemical composition of stars and interstellar clouds \cite{wakelam2017h2}; study of the physics of fusion and high-temperature plasmas \cite{fantz2021negative}. The single-electron detachment energy of the hydride ion in quantum chemistry can provide valuable information about the electronic structure and bonding of molecules, which are crucial for understanding chemical reactions and designing new materials. In addition, the reaction energy is an important descriptor to predict the outcomes of chemical reactions, optimize reaction conditions, and develop new chemical processes. For example, the water reacting with hydride ion is thought to be an important process in the interstellar medium \cite{herbst2021unusual}. Hydride ions are commonly used as reducing agents, and the reaction of water with hydride ion can lead to the production of hydrogen gas and hydroxide ions, which can be used in a variety of chemical reactions \cite{mcneill2020h}. By calculating potential energy surface and reaction energies, Calvin et. al. \cite{ritchie1968theoretical1, ritchie1968theoretical2} have demonstrated the two proton transfer reactions of hydride ion with water to form hydroxide and hydrogen molecule, and with hydrogen fluoride to form fluoride ion and hydrogen molecule.
For both the reactions they have calculated the energies with the LCAO-MO-SCF method using GTO's($H_{3}^{-}$ and $H_{2}F^{-}$system with both small and large basis type) and found the energy to be endothermic \cite{ritchie1967gaussian}. However, when they take the experimental data for all the molecules separately and calculate the reaction energy it comes out to be exothermic in nature.\cite{ritchie1968theoretical1,ritchie1968theoretical2}

In this paper, we have used VQE to calculate the ground state energy and single-electron detachment energy of hydride ion along with reaction energy of two-proton transfer reaction using IBM quantum machine. The paper is organized as follows: In theory and methodology, the construction of electronic Hamiltonian and the theory of the VQE process by explaining the classical ansatz (CC) as well as the quantum computational ansatz (q-UCC and HEA ansatz ) used in VQE is given. In the result and discussion part, we have calculated the minimum energy state and single-electron detachment energy of hydride ion through classical and quantum ansatz. We have studied the accuracy and relative error of our calculation with different algorithms. In the end, we calculated the single-electron detachment energy and reaction energy of the two proton transfer reactions and studied the accuracy of quantum computation.

\section{Theory and Methodology}

The description of any quantum mechanical system is given by writing the many-body Hamiltonian. The electronic Hamiltonian after applying Born-Oppenheimer approximation \cite{combes1981born} and in terms of second quantization can be represented as:

\begin{equation}
\label{eq:second quant}
H=\sum_{i, j} h_{i j} \beta_i^{\dagger} \beta_j+\frac{1}{2} \sum_{i, j, k, l} h_{i j k l} \beta_i^{\dagger} \beta_j^{\dagger} \beta_k \beta_l
\end{equation}

\noindent where $h_{i,j}$ and $h_{i,j,k,l}$ are the one and two electron integrals:

\begin{equation}
h_{i j}=\int d \overrightarrow{r_1} \chi_i^*\left(\overrightarrow{r_1}\right)\left(-\frac{1}{2} \nabla_1^2-\sum_\sigma \frac{z}{\left|\vec{r}_1-R_\sigma\right|}\right) \chi_j\left(\overrightarrow{r_1}\right)
\label{eq: two body integral 1}
\end{equation}

\begin{equation}
h_{i j k l}=\int d \overrightarrow{r_1 }\ d\overrightarrow{r_2 }\frac{\chi_i^*\left(\overrightarrow{r_2}\right)\chi_j^*\left(\overrightarrow{r_2}\right)
\chi_k^*\left(\overrightarrow{r_2}\right)\chi_l^*\left(\overrightarrow{r_2}\right)}{\left | r_{1}-r_{2} \right |}
\label{eq:two body integral 2}
\end{equation}

where ${\chi _{i}}^{*}\left(\overrightarrow{r_2}\right)$ is the $i^{th}$  spin-orbital, Z is the nuclear charge, $r_i$ is the $i^{th}$ electron's position, $r_{12}$ is the separation between two electrons, and $R_{\sigma }$ is the nucleus' position.The one-body integral depicts the kinetic energy of electrons and their interaction with nuclei, whereas the two-body integral represents the interactions between electrons. The numerical integration of these integrals is carried out using the Open fermion software \cite{mcclean2020openfermion}. The fermionic creation and annihilation operators, $\beta _{i}^{+}$ and $\beta _{j}$, raise and lowers an orbital's occupational number by one, respectively. For the second quantization, the anticommutation of these operators guarantees the antisymmetric nature: 
\begin{equation}
\left \{ \beta _{i}^{+}, \beta _{j} \right \} = \delta _{i,j}
\end{equation}
\begin{equation}
\left \{ \beta _{i}, \beta _{j} \right \} = 0
\end{equation}

Using the second quantized Hamiltonian, one can calculate the ground state energy using VQE, a hybrid algorithm developed by Perruzo et al \cite{peruzzo2014variational}, which is based on the concept of the Rayleigh-Ritz  variational principle \cite{gould2012variational},
i.e., the expectation value of a Hamiltonian is always greater than or equal to its smallest eigenvalue $E_{g}$, for any wavefunction $\ket{\psi(\theta)}$, where $\theta _{0}=(\theta _{0},\theta _{1},\theta _{2}\cdots \theta _{i})$
$\theta _{i}$ is the parameter of $i^th$ excitation \cite{tilly2022variational}.
\subsection{\textbf{Classical Computational Ansatz}}

 For preparing the reasonable parameter-dependent state ($\ket{\psi(\theta)}$), classically the best available method is Hartree-Fock(HF), Moller-Plesset (MP) perturbation theory \cite{moller1934note,leininger2000mo}, full configuration interaction(FCI), and coupled-cluster with single and double excitation(CCSD). 

The Slater determinant generated from the Hartree-Fock calculation is taken as a reference for the post-Hartree-Fock methods i.e. FCI and CCSD. FCI method expands the wavefunction in $\eta$-fock space but becomes quickly intractable due to a number of determinants \cite{sherrill1999configuration} which grows factorially with the number of electrons and orbitals. Therefore, one can take the Hartree-Fock state as a reference state to fix that problem \cite{sherrill1999configuration} and the excitation operators can be defined as: 

 \begin{equation}
T=\sum_{i=1}^{\eta} \hat{T}_{i}
 \end{equation}
 \begin{equation}
 \hat{T}_{1}=   \sum\limits_{\substack{i \in occ \\ a\in virt}} t_{a}^{i} a_{a}^{\dagger}a_{i}
 \end{equation}
\begin{equation}
\hat{T}_{2}=   \sum\limits_{\substack{i> j \in occ \\ a> b\in virt}} t_{ab}^{ij} a_{a}^{\dagger} a_{b}^{\dagger} a_{i}a_{j}
\end{equation}

\noindent Here the \textit{occ} and \textit{virt} represent the occupied and unoccupied sites in the reference state. The operator $\hat{T}_{1} $ and $\hat{T}_{2}$ generate the single excitation and double excitation, respectively, from the reference state. $t_{a}^{i}$ and $t_{ab}^{ij}$ are the expansion coefficients. The full wavefunction can be written as: 
\begin{equation}
|\Psi_{FCI}\rangle =(1+T)|\Psi_{HF}\rangle
\end{equation}
However, this method converges slowly for a highly correlated state. This problem is solved by the coupled cluster (CC) method which constructs the wavefunction in the exponential form as:
\begin{equation}
    |\Psi_{CC}\rangle =e^{T}|\Psi_{HF}\rangle
\end{equation}
 CC is a post-Hartree–Fock method that aims at recovering a portion of electron correlation energy by evolving an initial wave function (usually the Hartree–Fock wave function) under the action of parameterized excitation operators resulting in CC with single and double excitations (CCSD) represented as: 

 \begin{equation}
     |\Psi_{CCSD}\rangle =e^{\hat{T}_{1}+\hat{T}_{2}}|\Psi_{HF}\rangle
 \end{equation}
 In the coupled cluster formulation the operator $e^{T}$ is not unitary and the energy obtained is not variational in nature. Therefore, to implement this formulation on a quantum computer we need a unitary version of this operator which can be implemented using a unitary coupled cluster(UCC) ansatz.

\subsection{\textbf{Quantum Computational Ansatz}}
\begin{figure}[h]
 \begin{center}
 \includegraphics[scale=.65]{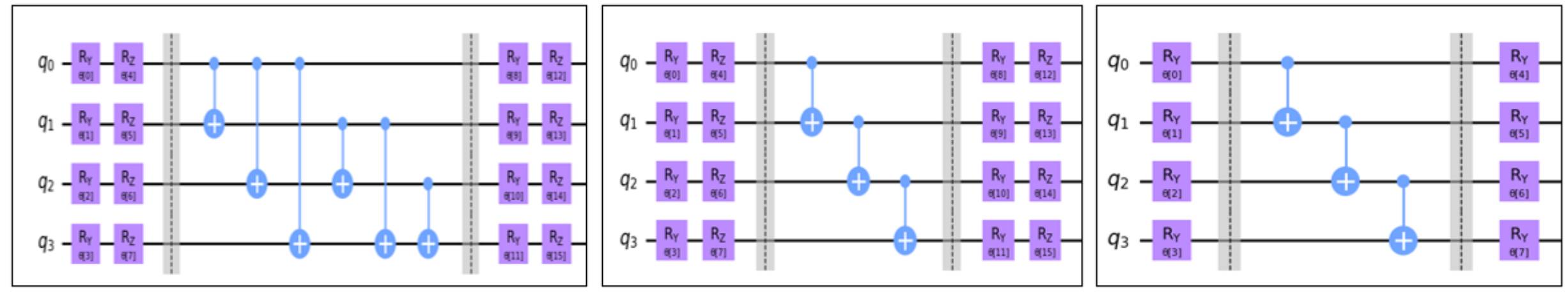}
\caption{Representation of one repetition of HEA in Qiskit, using Ry, Rz, and CNOT gates, on four qubits with full entanglement(RY-RZ-CX full ansatz) (left), RY-RZ-CX with linear entanglement (middle) and RY-CX ansatz  with full entanglement(right)}
\label{fig: HEA circuit}
\end{center}
\end{figure}

To implement a given quantum mechanical problem on quantum computers, generally, two categories of ansatz are used i.e. problem agnostic ansatz and problem tailored ansatz \cite{ramoa2022ans}. The problem agnostic ansatz is constructed from a limited set of quantum gates which are easy to implement on the quantum hardware, however, a chemical interpretation of different terms is generally not possible. They are also called hardware efficient ansatz (HEA) \cite{kandala2017hardware,tilly2021reduced}.  HEA usually employs layers of parameterized single-qubit gates, which can provide linear transformation, and layers of two-qubit gates, which can bring entanglement in the state and enrich the expressive power.
To show the accuracy of hardware efficient ansatz, we have taken three types of HEA (i) RY-RZ-CX ansatz with linear entanglement  (ii) RY-RZ-CX ansatz with full entanglement (iii) RY-CX ansatz with full entanglement as represented in 
Fig. \ref{fig: HEA circuit}

HEA can be constructed differently depending on the depth of the circuit, the type of rotation gates used, and the entanglement setting. HEA is made of a series of repeating rotation gates followed by CNOT gates. This is implemented using the EfficientSU2 function \cite{seu2} that creates a circuit with a specific pattern of gates. 

\begin{figure}[h]
 \begin{center}
 \includegraphics[scale=.4]{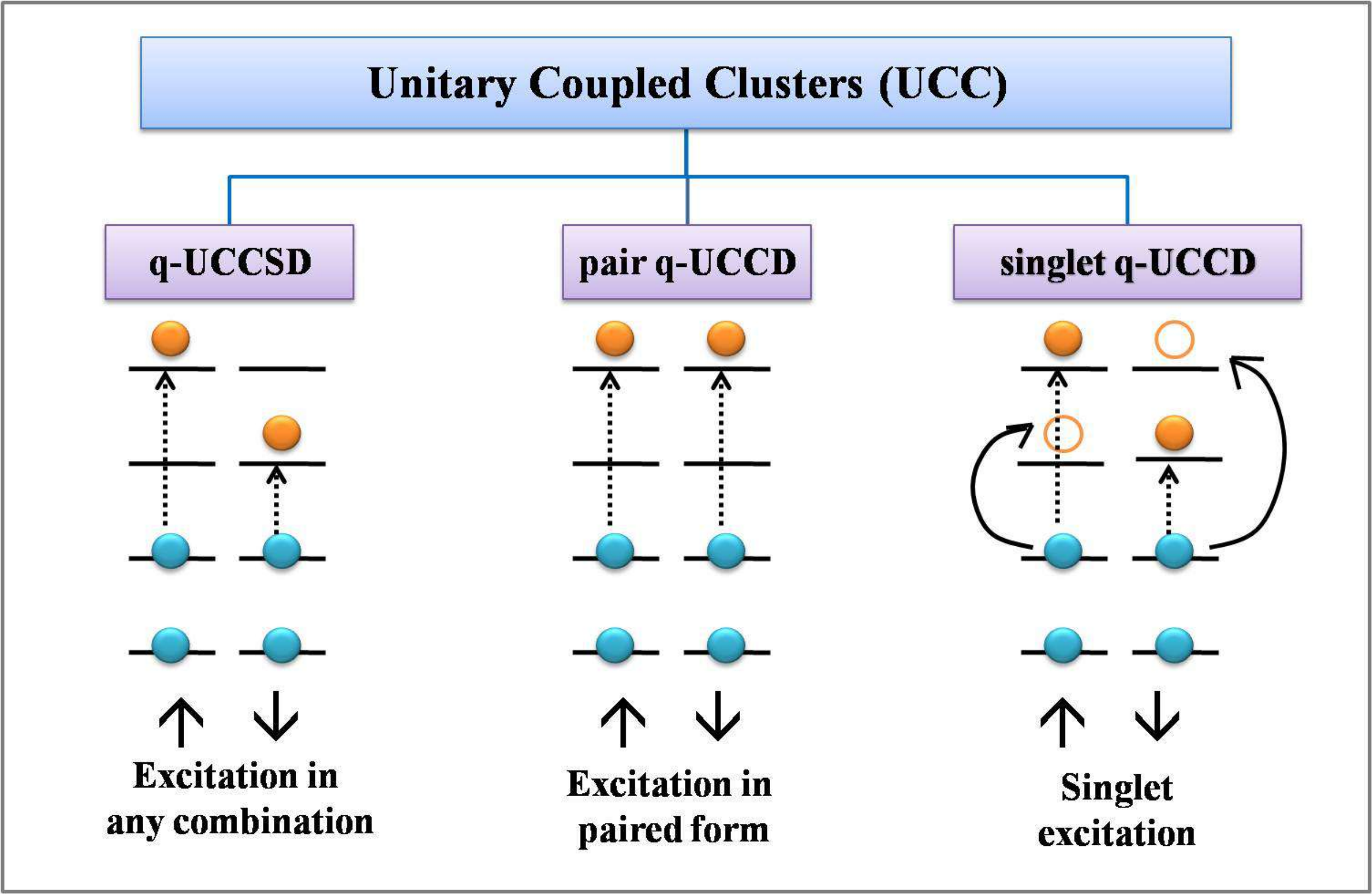}
\caption{Schematic diagram showing the three different types of quantum computational ansatz with their nature of excitation. singlet q-UCCD allows the singlet double excitation where the term "singlet" in singlet excited state refers to the fact that the electrons in the excited state have opposite spins. This means that the total spin of the excited state is zero. Similarly, pair q-UCCD allows excitation in the form of pairs (exciting two electrons from two paired occupied orbitals to two unpaired virtual orbitals) and q-UCCSD allows any combination of single and double excitation.}
\label{fig:ucc types}
\end{center}
\end{figure}

On the other hand, problem-tailored ansatz is chemistry-inspired ansatzes designed by using the domain knowledge from traditional quantum chemistry in a way that every term in the ansatz describes a certain electron configuration. 
These are the unitary version of the classical coupled cluster(UCC) \cite{romero2018strategies}. In this study, we have taken three types of chemistry-inspired ansatz (i) q-UCCSD ansatz(quantum-Unitary Coupled Cluster Single and Double excitation) \cite{uccsd},(ii) pair q-UCCD ansatz \cite{puccd} and (iii) singlet q-UCCD ansatz \cite{succd}, as represented in Fig. \ref{fig:ucc types}.

UCC ansatzes are based on Coupled Cluster Theory truncated at single and double excitations \cite{sokolov2020quantum}. 

In the unitary coupled cluster, the antisymmetric wavefunction is defined as:

\begin{equation}
    |\Psi_{UCC}\rangle =e^{T-T^{\dagger}}|\Psi_{HF}\rangle
\end{equation}

Since the UCC operator is unitary, the method is variational, and the UCC energy is an upper bound to the ground state energy. So, the energy expression can be written as 
\begin{equation}
    E_{UCC}= \min_{\underset{t}{\rightarrow}}^{}\langle \Psi_{HF}|e^{-(T-T^{\dagger})}He^{T-T^{\dagger}}|\Psi_{HF}\rangle
\end{equation}

The most popular UCC method is that where we only take the single and double excitation known as UCCSD. The wavefunction for the UCCSD ansatz become
\begin{equation}
\label{trotterize}
    |\Psi_{UCCSD}\rangle=e^{(T_{1}-T_{2})-(T_{1}^{\dagger}+T_{2}^{\dagger})}|\Psi_{HF}\rangle
\end{equation}

Note that UCC ansatz to operate on the quantum computer uses the trotterization method to write it in the form of the product of exponentials\cite{jones2012faster}, which allows recasting the expression as a quantum circuit with a sequence of the quantum gate.  The pair q-UCCD ansatz \cite{elfving2021simulating} allows only double excitation and the 'singlet q-UCCD ansatz' \cite{sokolov2020quantum} allows for singlet excitation.  Singlet q-UCCD splits the double excitations into singlet and triplet components and makes use of symmetry / anti-symmetry to reduce the number of excitation. Pair q-UCCD limits the excitations to those which excite a pair of electrons with opposite spins from one spatial orbital to another operator \cite{sokolov2020quantum}. The criteria for both the ansatz are that they work for the singlet spin systems i.e. their number of alpha and beta electrons have to be the same. The molecules that we have chosen for our study all belong to the singlet spin systems which gives us the advantage to check the variation in UCC ansatz.

\subsection{\textbf{Basis Sets }}
Basis sets are the crucial parameter that is important for the recovery of correlation energy.
Some of the most simple basis sets that are used in the classical computation are the STO-nG basis sets(Slater Type Orbital-n Gaussians)\cite{hehre1969self}.In the Slater-type orbital, each atomic orbital is considered to be an approximate STO. The STO-nG basis such as STO-3G and STO-6G is called the minimal basis sets as they contain the orbitals that are only needed for the Hartree-Fock calculation. Our study of hydride ion and other molecules involved in the proton transfer reaction needs better basis sets to recover the correlation energy. Therefore, we added the split valence basis functions\cite{ditchfield1971self} (3-21G,6-31G) which provides radial flexibility due to its double-zeta representation. Furthermore, to reach the experimental value through quantum computation we increased the basis size by taking  $631++G$,$6-311+G^{*}$, $631++G^{**}$ function which includes polarisation. However, the basis sets that do the job for hydride ion is the correlation-consistent basis sets which add additional virtual orbitals which help to recover the correlation energy \cite{dunning1989gaussian}. We took the aug-cc-pVDZ(polarized valence double-zeta) \cite{kendall1992electron} basis for hydride ion which took us near to the reference energy \cite{chandrasekhar1944some} but becomes an 18 qubits system for calculation. Since the number of qubits required for the computations is equal to the number of spin-orbitals (which is in turn decided by the choice of single-particle basis set), the qubit requirement for hydride ion is 4 for both 3-21G and 6-31G basis set. But for 6-31++G,$6-311+G^{*}$, $631++G^{**}$, and  aug-cc-pVDZ basis set increases to 6,6 and 12, respectively. This highlights the fact that working on a suitably large basis set is crucial for obtaining accurate results. In our study for the proton transfer reaction, we were only able to use the 3-21G, 6-31G, and 6-31++G basis sets beyond that the number of qubits needed for the system makes the process difficult.

 \subsection{\textbf{Total Energy Calculation with VQE }}

\begin{figure}[tbh]
 \centering
 
 \includegraphics[scale=.4]{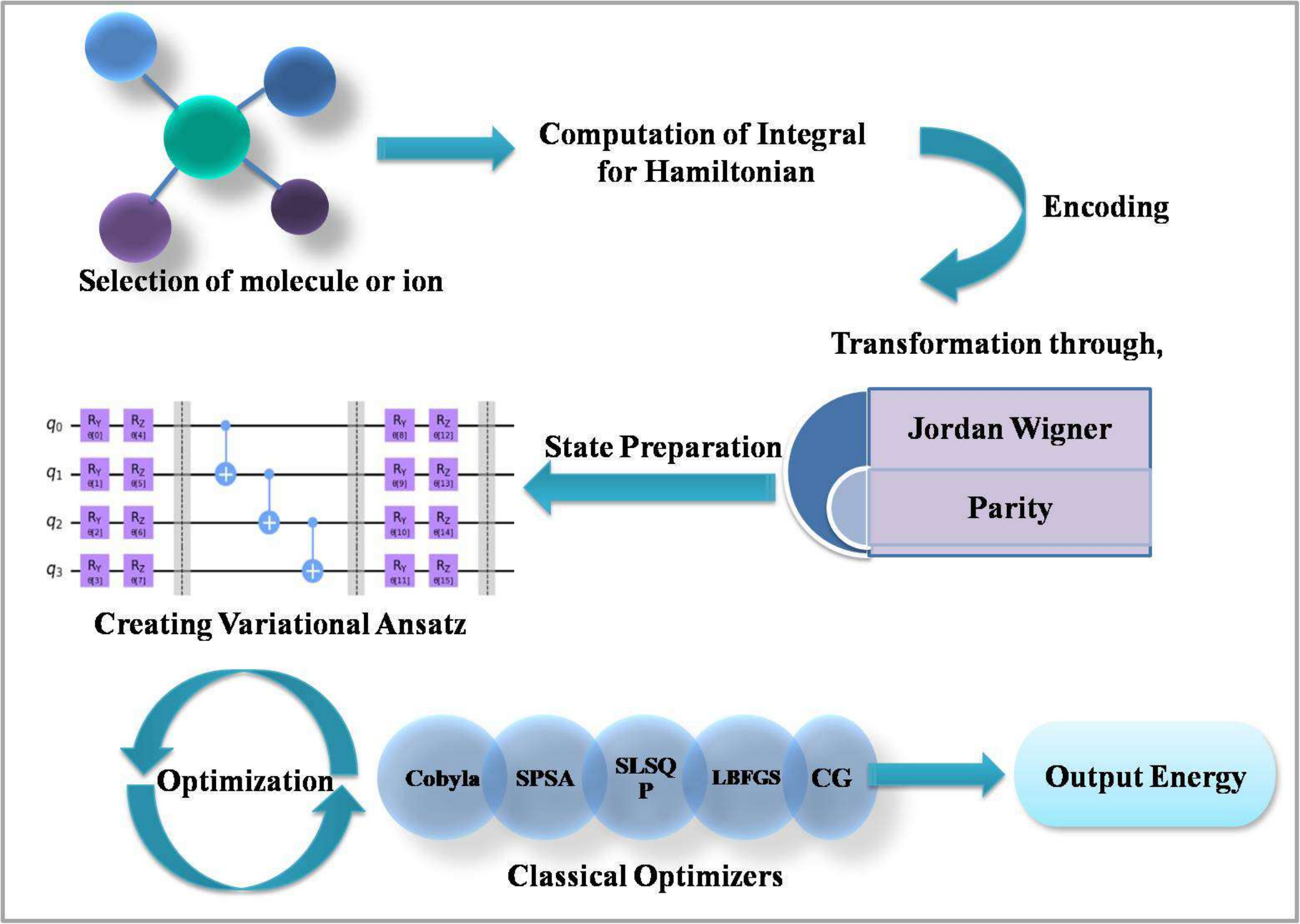}
\caption{A schematic representation of the VQE algorithm having the following steps: (i) Initialization of the molecule by providing the geometry, charge, and spin multiplicity. (ii) HF calculation by the PySCF package and the generation of fermionic Hamiltonian. (iii) Fermion to Pauli operator mapping by using mappers (JW or Parity). (iv) Providing the reference Hartree-Fock state to the ansatz and calculation of ground state energy. (v) Optimization process by using different optimizers such as COBYLA. (vi) Obtaining the  final VQE ground state energy which strictly follows the variational principle.}

\label{fig:vqe}
\end{figure}

To solve a fermionic problem using a quantum computer, it is necessary to encode the evolution of a fermionic state as a unitary operation on the state of the qubits (which are represented by Pauli operators). This requires mapping the fermionic operators to spin operators and subsequently to qubits. 
Mainly the operators are mapped into a string of tensor products of Pauli operators X, Y, and Z which can easily be measured. In our study, we have used the simplest Jordan- Wigner mapping procedure  \cite{jordan1993paulische} for the hydride ion simulation. In Jordan-Wigner's mapping, the number of qubits is equal to the number of spin orbitals\cite{}. However, for the calculation of bigger molecules like $H_{2}O$, we have used Parity mapping \cite{bravyi2017tapering}. The one advantage that parity mapping provides is tapering off the number of qubits. Usually, two qubits can be reduced by using parity mapping as it uses the parity basis with \ce{Z2} symmetry. After the fermionic to spin operator we then trotterize the UCC operator (Eq. \ref{trotterize}). Then the expectation value of the mapped Hamiltonian can be calculated using two methods,(i) State-Vector approach using Qiskit state-vector backend,(ii) Measurement approach using Qiskit QASM backend. To understand the important steps of the VQE we have used a graphical representation as shown in Fig \ref{fig:vqe}.

After calculating the energy through the Aer-simulator \cite{aer} provided by Qiskit\cite{qiskit} which uses a matrix representation of the operator in the Hilbert space, we pass the energy to a classical optimizer. The classical optimizer minimizes the energy that we have obtained from the VQE with respect to the obtained parameters. Again when the new parameters are obtained they work as input for the ansatz from the previous step. This process is repeated until the energy is minimized. VQE assures that the energy that we obtain from VQE is greater than the true energy. Therefore, choosing a good optimizer is another task. For our calculation, we have used the COBYLA optimizer which uses the Constrained Optimisation by Linear Approximation method\cite{powell1994direct}. The algorithm optimizes a constrained problem where the function has no derivative. For the HEA we have used the SPSA (Simultaneous Perturbation Stochastic Approximation) optimizer \cite{alteg2022study,spall2000adaptive}. Optimizers have a direct impact on the number of measurements needed to complete an optimization as well as on the number of iterations required to obtain convergence. For the hydride ion calculation, we have gone up to 1000 iterations but for the proton transfer reaction, we have extended it up to 3000 iterations to achieve chemical accuracy.

We carried out the initial or reference state (Hartree-Fock state), and the classical CCSD calculations using PySCF \cite{sun2020recent}, while the q-UCCSD-VQE computations were performed using the Qiskit program.  We used the latest Qiskit  $ 0.40.0$ package\cite{qiskit}  to carry out quantum simulations using the VQE algorithm. FCI energy was obtained by the exact diagonalization of  Hamiltonian in the given basis sets using the OpenFermion \cite{mcclean2020openfermion} package. To simulate our circuits/ansatzes, we use the AerSimulator backend \cite{} from IBM Qiskit \cite{anis2021qiskit} which works as a noise-free quantum device.

\section{Results and Discussion}

\subsection{\textbf{Ground State Energy Estimation}}

The non-relativistic ground state energy of hydride ion was previously calculated to be -0.5253 Ha by Hylleraas\cite{hylleraas1929neue} using his variational approach with the wavefunction:  

\begin{equation}
    \psi (r_{1},r_{2})=e^{-mr_{1}-nr_{2}}+e^{-nr_{1}-mr_{2}}
\end{equation}

\noindent where m and n are variational parameters representing the effective nuclear charges of the electrons. The energy was corrected to -0.52592 Ha by Chandrasekhar\cite{chandrasekhar1944some} using electron-electron correlation implicitly in the wave function as 
\begin{equation}
\label{eq:chandrasekhar}
    \psi =\psi (r_{1},r_{2})(1+kr_{12})
\end{equation}
where k is the new variational parameters. 
The stability of the bound state of hydride ion was in question that has been asked for many years. The nature of instability makes recovering the correlation energy difficult in the post-Hartree Fock method. In general, for any correlated method (FCI, CCSD,
q-UCCSD, singlet-UCCD and pair q-UCCD, etc.) the total exact ground state energy can be decomposed into an HF (mean-field) and a correlation contribution as:
\begin{equation}
   E_{exact} = E_{HF} + E_{correlation}
   \implies E_{correlation} = E_{exact} - E_{HF}
   \label{eq:correlation equation}
\end{equation}

where $E_{exact}$ represents the energy calculated by different methods to recover that correlation energy. Therefore, the method which recovers most of the correlation energy of a system is considered as the accurate method. Table \ref{tbl:hydride ion table} shows the energy obtained using the different classical and quantum computational ansatz.

The main components that both ansatz (chemistry-inspired and HEA) depends on are the number of parameters and the depth of their circuit. Increasing the number of parameters in the ansatz allows for a larger representation of the Hilbert space, potentially leading to more accurate approximations of the ground-state energy, while increasing the depth of the circuit  (number of CNOT gates) can improve the quality of the ansatz state, but also increase the probability of encountering noise and errors. Finding the right balance between these factors is crucial for achieving accurate and efficient VQE calculations.
To verify this we consider the hydride ion ground state calculation.

\begin{figure}[H]
\centering

\includegraphics[scale=.15]{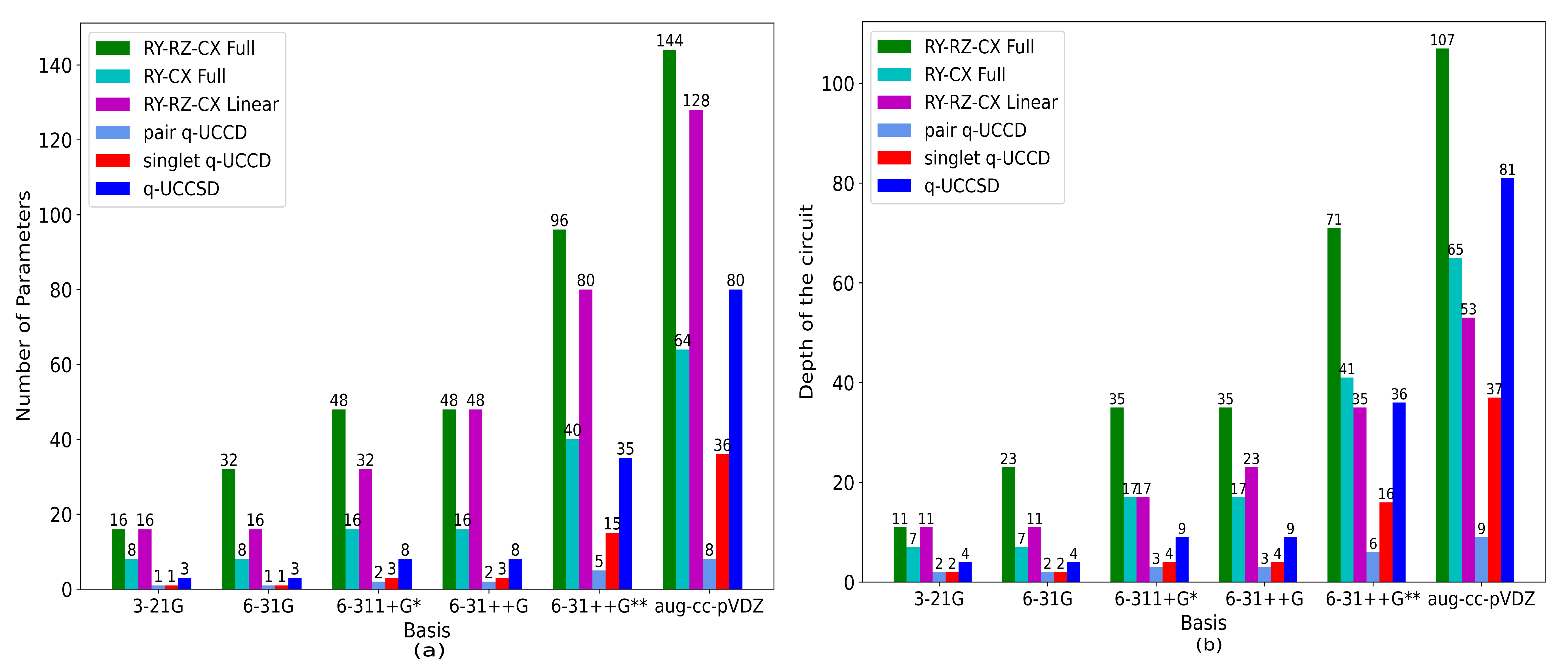}
\caption{(a) The number of parameters used by the hardware efficient ansatz(RY-RZ-CX linear ansatz, RY-RZ-CX full entanglement and RY-CX full entanglement ansatz) and chemistry inspired ansatz ( singlet q-UCCD, pair q-UCCD and q-UCCSD) for the hydride ion ground state energy calculation in combination with different basis sets. (b) The depth of the circuit (Number of CNOT gates used) for the different hardware efficient ansatz (RY-RZ-CX linear ansatz, RY-RZ-CX full entanglement, and RY-CX full entanglement ansatz) and Chemistry inspired ansatz with increasing size of basis sets.}
\label{fig:abc}
\end{figure}

It is evident from Fig. \ref{fig:abc} that, the parameters and depth increase with respect to the choice of the ansatz and the basis sets. It should be kept in mind that an increase in the size of basis sets is actually proportional to an increase in the number of qubits, as each extra orbital in the basis adds one extra qubit to the system. 
 We can see from  Fig. \ref{fig:abc} that more parameters and depth are being used by the HEA for the hydride ion case.
If we consider the largest basis in our study (aug-cc-pVDZ) which takes 18 qubits for the simulation of a hydride ion, the number of parameters used by the HEA such as RY-RZ-CX with linear entanglement, RY-RZ-CX with full entanglement and RY-CX with full entanglement takes 144, 128 and 64, respectively. While the number of parameters used with chemistry-inspired ansatz i.e. q-UCCSD, singlet q-UCCD, and pair q-UCCD takes 80,36 and 8 parameters, respectively. The number of parameters used by chemistry-inspired ansatz is lower than the HEA. The main issue in the HEA ansatz is the increasing depth of the circuit. For the hydride ion, the maximum number of CNOT gates is used by the RY-CZ-CX with linear entanglement ansatz, and the lowest is used by the RY-CX ansatz. Similarly, the highest and lowest number of parameters and depth used by chemistry-inspired ansatz is the pair q-UCCD ansatz.

\begin{table}[H]
  \caption{The energy obtained for the hydride ion from the classical computational ansatz ( HF, FCI, CCSD ) and quantum computational ansatz (singlet q-UCCD, pair q-UCCD and q-UCCSD) with respect to different basis sets. All the energies calculated are in units of Hartree (Ha). The reference energy for the total energy of hydride ion is taken with respect to Chandrasekhar wavefunction i.e.-0.52592 Ha. \cite{chandrasekhar1944some}}
  \label{tbl:hydride ion table}
  \resizebox{\textwidth}{!}{
  \begin{tabular}{|c|c|c|c|c|c|c|}
    \hline
    Basis  & Hartree-Fock (HF) & FCI  &  CCSD & singlet q-UCCD  & pair q-UCCD & q-UCCSD\\
    & Energy & Energy & Energy & & & \\
    
    \hline
    
      3-21G & -0.40042019 & -0.40850110 & -0.40042019 & -0.40843457 & -0.40843457 & -0.40850109\\
    \hline
    
    6-31G & -0.42244193 & -0.43137772 & -0.43137771 & -0.43129428 & -0.43129428 & -0.43137771\\
    
    \hline
      6-311+$G^{*}$ & -0.46667188 & -0.48163236 & -0.48163200 & -0.48051852 & -0.47952746 & -0.48163212 \\
     \hline
     6-31++G & -0.48707258 & -0.51001427  & -0.51001422 & -0.50575549 &  -0.50243001 & -0.51001421\\
     \hline 
     6-31++$G^{**}$ & -0.48707258 &  -0.51238663 & -0.51238659 & -0.50575549  & -0.50243001 & -0.51238663\\
    \hline
    aug-cc-pVDZ & -0.48678004 & -0.52402862 & -0.52402900  &  -0.51813663 & -0.51254604 & -0.52402845\\
    \hline
  \end{tabular}
  }
\end{table}

Once the ansatz circuit is constructed for the unstable hydride ion by keeping the eye on the depth, it is important to study the outcomes of the ground state. For the basis sets 3-21G and 6-31G, the Hartree-Fock  total energy obtained for the hydride ion  is -0.40042019 Ha and -0.42244193 Ha, respectively which underestimates the reference value ground state energy of -0.52592 Ha\cite{chandrasekhar1944some}.

Using the best basis set for our calculation ( aug-cc-pVDZ) the Hartree-Fock energy is obtained as -0.48678004 Ha. To obtain the accuracy of reference energy value \cite{chandrasekhar1944some}, our first post-Hartree- Fock method FCI converges up to -0.52402862 Ha.
In comparison, the CCSD method converges at the value of -0.52402900 Ha. The verification of the bound state is confirmed from the 6-31++G basis when the ground state energy of the hydride ion becomes smaller than $\sim -0.5$ Ha. The nearest result up to the reference energy (-0.52592 Ha) is given by the correlation consistent basis set aug-cc-pVDZ. 
Taking the quantum computational ansatz into account, although the depth and parameters used by singlet q-UCCD ansatz and pair q-UCCD ansatz are less, they converged up to -0.51813663 Ha and -0.51254604 Ha. Both the ansatz failed to mimic the result of classical computational energy as well as the reference energy. On the other side, the q-UCCSD ansatz  converged up to  -0.52402845 Ha for the 18 qubit system (aug-cc-pVDZ basis) and shows a good agreement with the CCSD energy and the FCI energy.

\begin{figure}[H]
 \begin{center}

 \includegraphics[scale=.15]{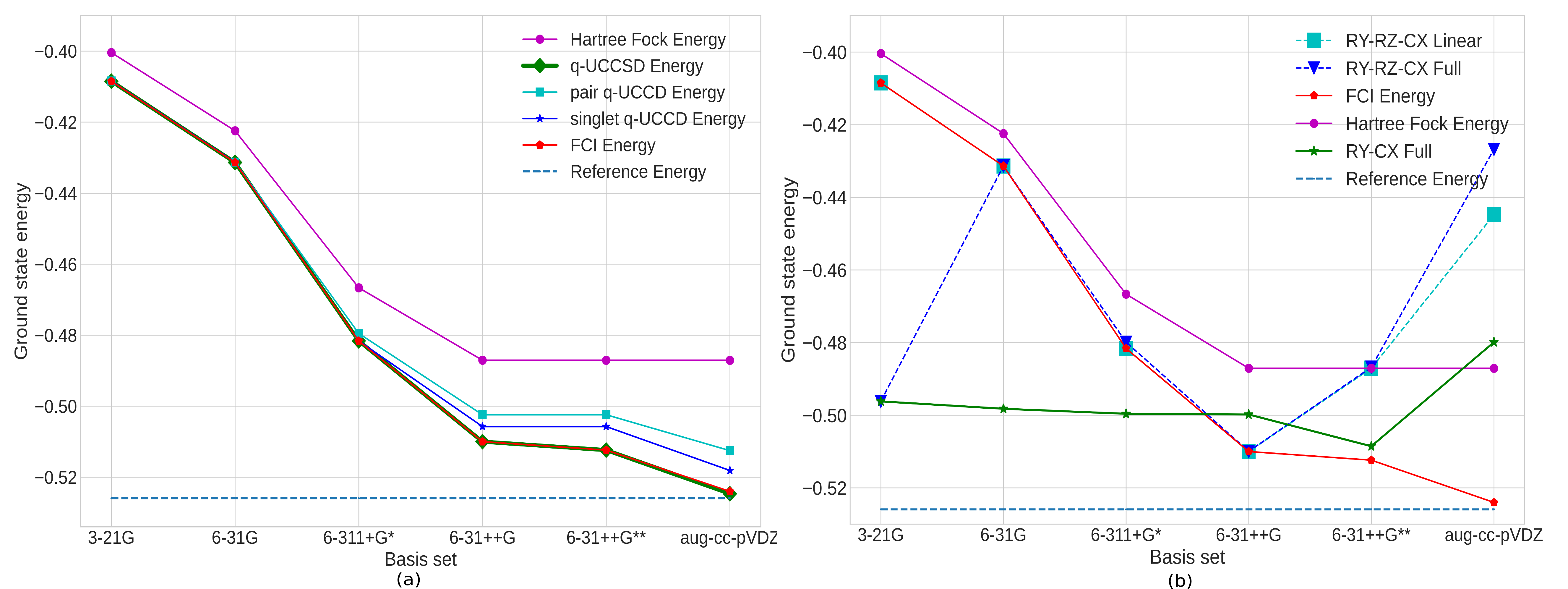}
\caption{(a) The variation in ground state energy of hydride ion with respect to the increasing number of qubits( i.e. change in basis sets). The number of orbitals used in a basis set is equal to the number of qubits for the calculation. The calculations are done for the chemistry-inspired ansatz( q-UCCSD, singlet q-UCCD, pair q-UCCD). (b) The variation in ground state energy of hydride ion with respect to the increasing number of qubits using the HEA (RY-RZ-CX linear entanglement, RY-RZ-CX with full entanglement, and RY-CX with full entanglement). The energies obtained are compared with the classical computational ansatz FCI.  The blue dotted line is showing the reference energy value for the hydride ion (-0.52592 Ha).  All energies in the y-axis are in the units of Hartree (Ha)}.
\label{fig:Chemsitry inspired and hardware efficient energy graph}
\end{center}

\end{figure}

Table \ref{tbl:h12} shows the ground state energy obtained through HEA (RY-RZ-CX with linear entanglement, RY-RZ-CX with full entanglement, and RY-CX with full entanglement). In Table \ref{tbl:h12}, the energies obtained from HEA are only compared with the Hartree-Fock energy and the FCI energy as for the hydride ion case both classical ansatzes (CCSD and FCI) show almost equal energy values. We can see with the starting basis (3-21G, 6-31G basis) the energy obtained for hydride ion  with the HEA (RY-CX with full entanglement) is far better as compared to the energy computed with the quantum computational ansatz and FCI method.  
With random initial reference states, the HEA is converging to poor values, therefore, we have provided the Hartree-Fock reference state to these calculations. As the number of qubits is increasing the circuit depth (CNOT gate) for the ansatz is increasing, which is the main reason for not converging the energy at higher basis sets. For hydride ion energy calculation we conclude the RY-CX ansatz to be the most reliable ansatz among the other HEA (RY-RZ-CX with linear entanglement, RY-RZ-CX with full entanglement)  (Table \ref{tbl:h12}).

\begin{table}[H]
\resizebox{\textwidth}{!}{
  \caption{Ground state energy of hydride ion with quantum computational ansatz based on hardware efficient method. All the energies calculated are in units of Hartree (Ha)}
  \label{tbl:h12}
  \begin{tabular}{|c|c|c|c|c|c|c|}
  
    \hline
    Basis  & Hartree-Fock (HF) & FCI Energy &  RY-RZ-CX & RY-RZ-CX  & RY-CX \\
    & Energy & & (Linear)& (Full)& (Full)\\
    
    \hline
    
      3-21G & -0.40042019 & -0.40850110 & -0.40850110 & -0.49619855 & -0.49619848  \\
    \hline
    
    6-31G & -0.42244193 & -0.43137772 & -0.43137772 & -0.43137772 & -0.49823275\\
    
    \hline
      6-311+$G^{*}$ & -0.46667188 & -0.48163236 & -0.48160259 & -0.47989550 & -0.49961421 \\
     \hline
     6-31++G & -0.48707258 & -0.51001427  & -0.51001359 & -0.50996433 & -0.49980979 \\
     \hline 
     6-31++$G^{**}$ & -0.48707258 & -0.51238663 & -0.48707242 &  -0.48678026 & -0.50855540  \\
    \hline
    aug-cc-pVDZ & -0.48678004 & -0.52402862 & -0.44478902  & -0.42678902 & -0.47989550  \\
    \hline
  \end{tabular}
  }

\end{table}

Even though the RY ansatz has
fewer costly CNOT gates than others, it does not converge to the accurate value of the ground state. This is due to the fact that at this low depth, the RY ansatz does not have sufficient coverage of the target Hilbert space. Increasing the depth of the RY ansatz will increase its coverage of the target space, but also increases the number of CNOT gates. This is not an equivalent trade-off and actually decreases the ability of the ansatz to approximate the ground state. We have used Fig.\ref{fig:Chemsitry inspired and hardware efficient energy graph} to correctly observe the accuracy trend of HEA and chemistry-inspired ansatz for simulating the hydride ion with the VQE method. Although the RY-RZ-CX full and RY-RZ-CX- linear entanglement ansatz uses more parameters to cover the Hilbert space but higher depth of the circuit puts them in a difficult situation to converge to the true ground state.

\begin{figure}[tbh]
 \begin{center}
 
 \includegraphics[scale=.16]{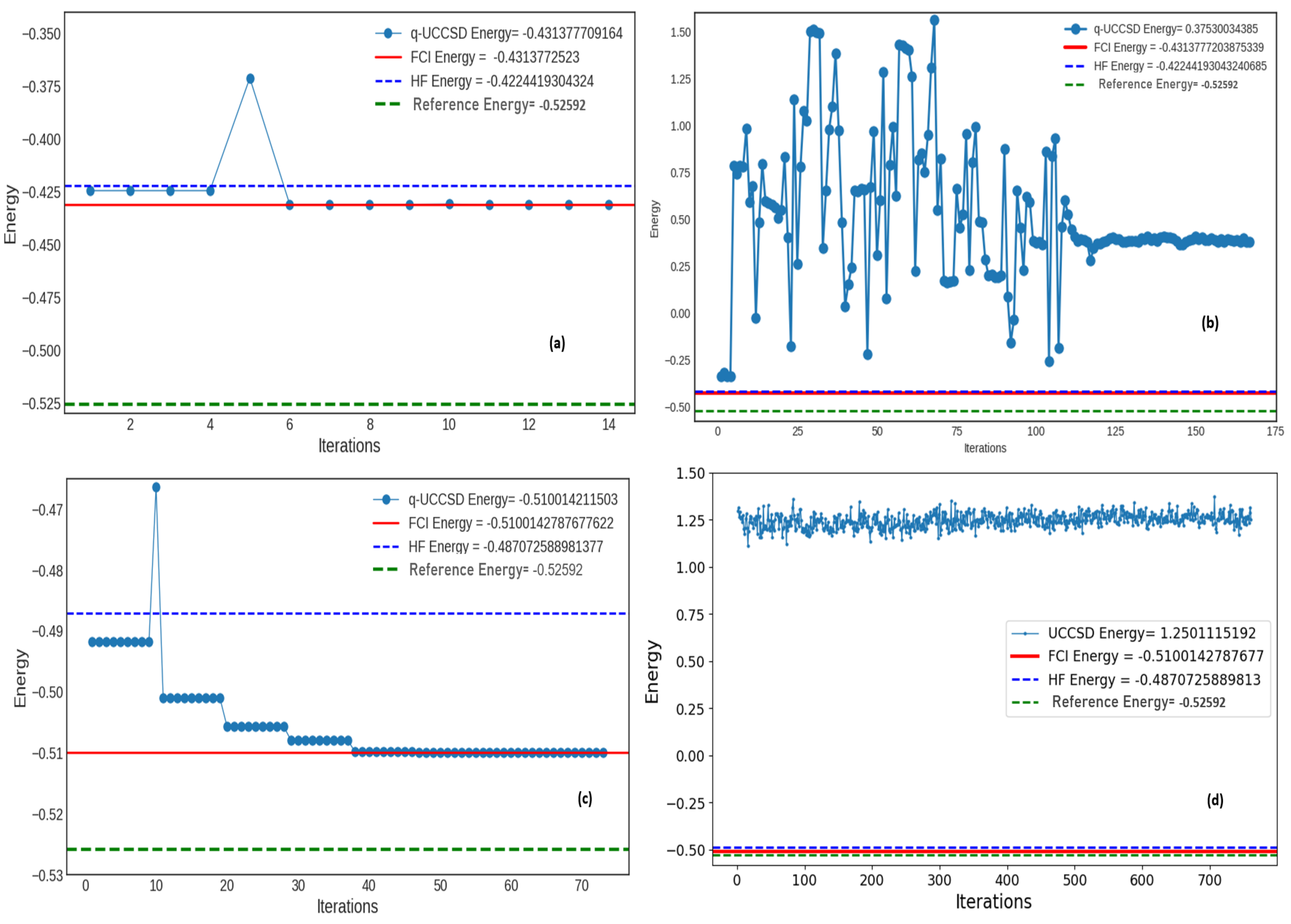}
\caption{Ground state energy as a function of the number of iterations for hydride ion. (a) and (b) are for the noiseless quantum computer(Aer simulator) and noisy quantum computer(ibmq-lima for the 4-qubit quantum simulation for 6-31G basis set. (c) and (d) are for the noiseless quantum computer (Aer simulator) and noisy quantum computer (ibmq-nairobi for the 6-qubit) simulation using 6-31++G basis set. All energies in the y-axis are in the units of Hartree (Ha)}.
\label{fig:simulator and original quantum computer}
\end{center}
\end{figure}

Qiskit provides simulators as well as real quantum hardware to simulate the results. For hydride ion, we have gone up to 18 qubits to get the accuracy in ground state energy, which is difficult to run in a real quantum computer as till now Qiskit \cite{qiskit} only provides 7 qubits access to real quantum computers to its users. It gave us the opportunity to test the behavior of real hardware by using some basis sets (6-31G basis and 6-31++G) and check whether with a noisy environment, is real hardware able to get the accuracy with respect to the FCI energy. Fig. \ref{fig:simulator and original quantum computer} describes the behavior of the noiseless quantum computer and the noisy quantum computer for hydride ion. We have taken the Aer simulator as a noiseless quantum computer and ibmq-lima (access to 5 qubits) and ibmq-nairobi (access to 7 qubits) for our study. For 6-31G basis hydride ion only uses qubits about 4 and the FCI energy is about -0.431377225 Ha, the q-UCCSD ansatz is easily able to match the energy with a noiseless quantum computer. The original hardware (Fig. \ref{fig:simulator and original quantum computer} (b)) was unable to recover the correlation and was not even close to the FCI energy. Similarly, basis 631++G uses (6 orbitals and 6 qubits) for simulation of hydride ion and recovers correlation energy upto ~ - 22.9416 mHa both with respect to CCSD and FCI energy, shows accurate behavior with Aer simulator but fails to match the behavior with original hardware (ibmq-nairobi) as shown in Fig. \ref{fig:simulator and original quantum computer} (c) and (d)

The relative percentage error of the q-UCCSD method with respect to the FCI calculation with the aug-cc-pVDZ basis system is ~ $10^{-5}$. Similarly, for the singlet q-UCCD and the pair q-UCCD, the relative percentage error is ~ $10^{-1}$. We can conclude that the singlet hydride ion state provides better accuracy with single and double excitation ansatz from the singlet excitation and pair excitation. We can conclude from the graph that the correlation consistent basis set (aug-cc-pVDZ) using 18 orbitals provides the best result for the hydride ion ground state.

\begin{figure}[tbh]
 \begin{center}
 \includegraphics[scale=.13]{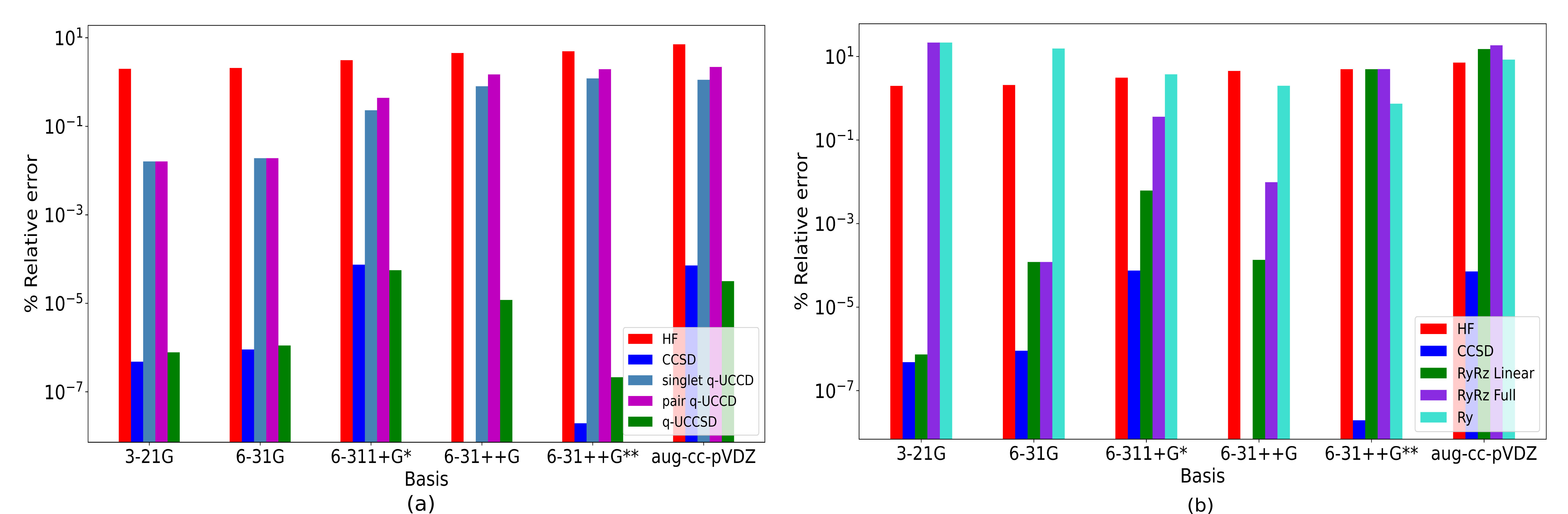}
\caption{The percentage relative error with respect to FCI for the ground state energies of the hydride ion both with respect to chemistry inspired ansatz and HEA, across six basis sets}
\label{fig:relative error hardware and chemistry}
\end{center}
\end{figure}

Fig.\ref{fig:relative error hardware and chemistry} presents the percentage relative error in ground state energy of hydride ion across the chosen basis sets for both chemistry-inspired ansatz and HEA. The correction to the correlation energy
increases with the number of simulated qubits (see Fig. \ref{fig:relative error hardware and chemistry}), while the number of required two-qubit gates also increases. Furthermore, it can be seen that for a fixed, large number of qubits (e.g., 18 available qubits for simulating the hydride ion) a larger portion of the correlation energy can be captured as compared to when applying to a smaller basis set.  
From Fig.\ref{fig:relative error hardware and chemistry} we can see the FCI energy and classical CCSD energy are the same in most of the cases for the hydride ion. The q-UCCSD has the lowest error in comparison to the other quantum computational ansatz.
The low number of function evaluations of the RY-CX HEA confirms that the simplest ansatz containing the solution will also have the best performance because the solution space has fewer parameters to optimize.


\begin{table}[H]
  \caption{The correlation energy obtained using different methods. The reference correlation energy of hydride ion is obtained by subtracting HF energy from the reference value of hydride ion (-0.52592Ha) \cite{chandrasekhar1944some}.}

  \centering
  \resizebox{\textwidth}{!}{
  \begin{tabular}{|c|c|c|c|c|c|c|}
    \hline
    Basis  & $E_{corr}$ & $E_{corr}$ & $E_{corr} $ & $E_{corr} $  & $E_{corr} $ & $E_{corr} $\\

    & (Ref.) & (FCI) & (CCSD) & (singlet  q-UCCD) & (pair q-UCCD) & (q-UCCSD)\\
    
    \hline
    
      3-21G & -0.125499 & -0.038847 & -0.008080 & -0.008080 & -0.008014 & -0.008080\\
    \hline
    
    6-31G &  -0.103478 &  -0.008935& -0.008935 & -0.008852 & -0.008852 & -0.008935 \\
    
    \hline
      6-311+$G^{*}$ & -0.059248 & -0.014960 & -0.014960 & -0.013846 &  -0.012855& -0.012855 \\
     \hline
     6-31++G & -0.038847 & -0.022941  & -0.022941 & -0.018682 &  -0.015357 & -0.022941\\
     \hline 
     6-31++$G^{**}$ & -0.038847 & -0.025314  & -0.025314 & -0.018682  & -0.015357 & -0.025314\\
    \hline
    aug-cc-pVDZ & -0.039139 & -0.037248 & -0.037248  &  -0.031356 & -0.025766 & -0.037248\\
    \hline
  \end{tabular}
  }
  \label{tbl:correlation table}
\end{table}

Table \ref{tbl:correlation table} represents the data of the recovered correlation energy by the post-Hartree-Fock methods.
If we took the energy calculated through Chandrasekhar wavefunction (-0.52592 ha) as the reference value, the correlation we need to obtain for the hydride ion in aug-cc-pVDZ basis to match the reference value is -39.1399 mHa. Therefore, it concludes although CCSD is showing better results than FCI it is still $\sim 2$ mHa behind the required correlation.
It can be seen in Table \ref{tbl:correlation table}, for 3-21G,6-31G basis sets the correlation energy obtained  through all the three ansatzes (q-UCCSD, singlet q-UCCD, and pair q-UCCD) is nearly same $\sim -8.0143$ mHa. With each increase in the size of the basis sets and the increase in the number of qubits the correlation energy obtain also increases at a rate of $\sim 4 $mHa. However, as the system size increases the singlet q-UCCD and pair q-UCCD ansatz shows fluctuation in the ground state energy and recovers correlation energy of -31.356 mHa and -25.766 mHa. However, the unitary ansatz with single and double excitation (q-UCCSD) recovers the maximum correlation energy among the other unitary ansatz (singlet q-UCCD and pair q-UCCD). About - 37.248 mHa  of correlation energy is recovered which is equal to the correlation energy obtained through the CCSD method. It concludes that the VQE algorithm with q-UCCSD ansatz is perfectly able to mimic the behavior of the best classical ansatz (CCSD).

\subsection{Single-electron Detachment Energy of Hydride ion}

Classical computation offers several methods to calculate single-electron detachment energy (Ionization energy) \cite{sahoo2009relativistic}, but VQE has not been utilized to calculate the electron detachment energy of such a reactive ion which is  abundant in interstellar space. Using VQE-qUCC ansatz to determine the single-electron detachment of the hydride ion will provide insight into the accuracy of this method in recovering the correct electron detachment energy of the ion as  this energy is the difference in the ground state energy of single ionized and neutral atom species. The single-electron detachment energy of the hydride ion has been experimentally determined by Lykke et.al.\cite{lykke1991threshold,rau1996negative}.  They used coaxial laser-ion beam photodetachment apparatus in their experiment and found the single-electron detachment energy of hydride ion to be $6082.99\pm 0.15 cm^{-1}$($\sim$ 0.75 eV or $\sim$ 0.027561990 Ha) where 1$cm^{-1}$ is equal to 0.0000046 Hartree.

\begin{figure}[tbh]
 \begin{center}
 
 \includegraphics[scale=.45]{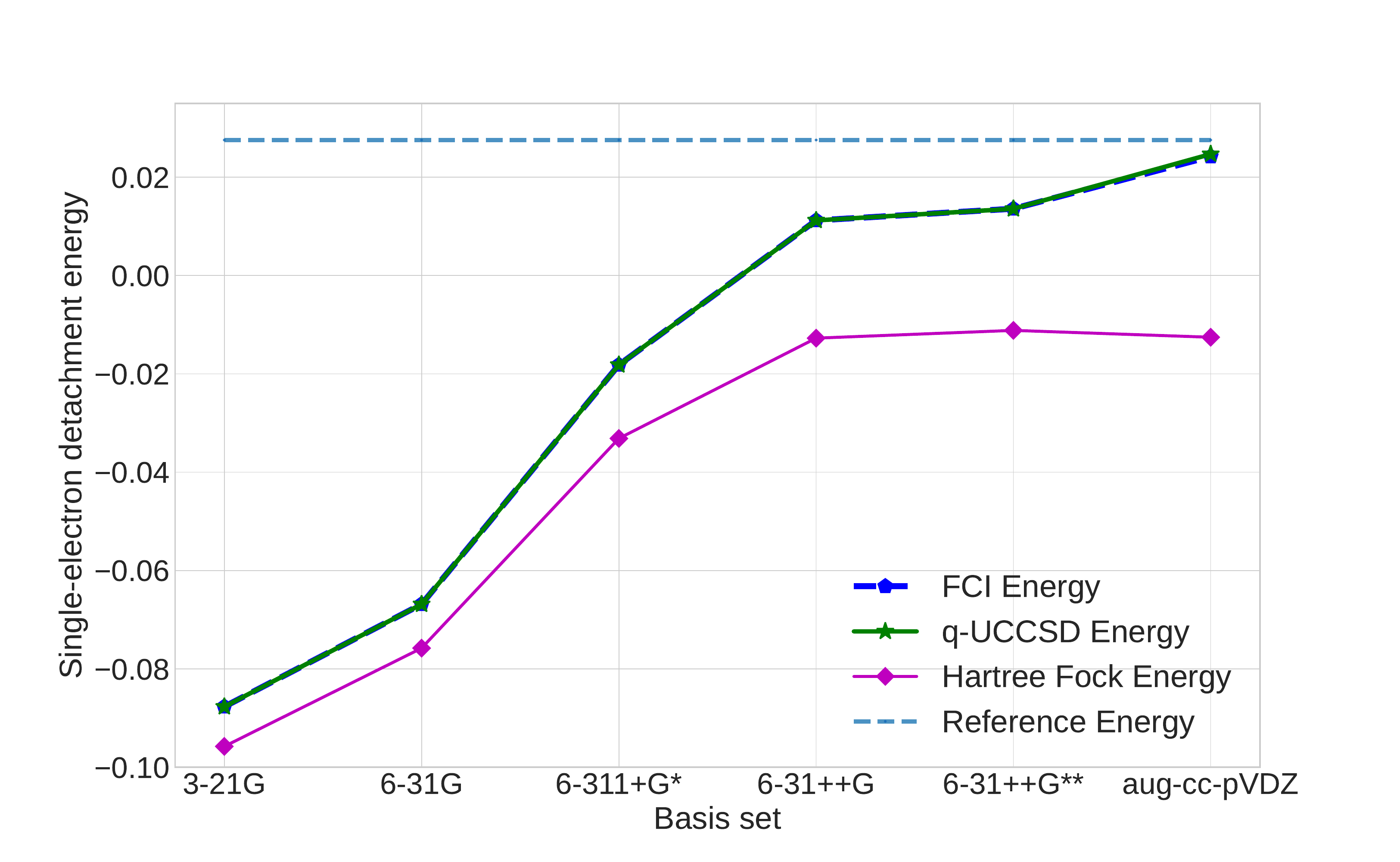}
\caption{Single-electron detachment energy of hydride ion calculated from the q-UCCSD, FCI, and HF method across six basis sets. The reference value of experimental energy  is ~ 0.749999 electron volt (0.027561990 Ha)\cite{lykke1991threshold}.  All energies in the y-axis are in the units of Hartree (Ha)}.
\label{fig: Ionization energy}
\end{center}
\end{figure}

 VQE algorithm needs to efficiently capture the right trends in the many-body effects in the ground state energies consistently for both the atom and corresponding ion. Calculating single-electron detachment energy using VQE serves as a qualitative and quantitative test of the algorithm. \cite{villela2022high}. For hydride ion, the energy obtained for a single-electron hydrogen atom (H) through the q-UCCSD method matches with the FCI value exactly. So taking the values of the hydride ion from the table (\ref{tbl:hydride ion table}). we calculated the single-electron detachment energy. Using q-UCCSD ansatz for aug-cc-pVDZ basis which 16 qubits for hydride ion, we calculated the single-electron detachment energy to 0.024694149 Ha (67 eV) which closely matches with the previously  calculated value $\sim$ 0.75 eV (0.027561990 Ha)\cite{rau1996negative,lykke1991threshold}. The variation of single-electron detachment energy for hydride ion is represented in Fig.\ref{fig: Ionization energy}.

\subsection{Proton Transfer Reaction Energy}
Reaction energy calculation is also a tedious task it will depend on all the molecule's behavior involved in the reaction mechanism. We have carried out the simulation of a set of molecules (${H_{2}}^{}$, ${OH_{}}^{-}$, \ce{H2O}, HF, ${H_{}}^{-}$, ${F_{}}^{-}$ ) involved in the two proton transfer reaction: 

\begin{equation}
    H^{-}+H_{2}O\rightarrow OH^{-}+H_{2}
    \label{eq:proton transfer reaction 1}
\end{equation}

\begin{equation}
    H^{-}+HF\rightarrow F^{-}+H_{2}
     \label{eq:proton transfer reaction 2}
\end{equation}

In Table \ref{tbl: number of orbital and parameters}, we have shown the number of parameters in q-UCCSD ansatz as a function of the number of orbitals being used. For any system, the higher the number of parameters more is the coverage of Hilbert space but difficult in simulation. A large number of amplitudes lengthen circuits, placing more work on the quantum computer. At the same time, more parameters must be optimized inside the traditional VQE loop.

\begin{table}[h]
  \caption{The number of orbitals and the number of parameters required for the VQE calculation using q-UCCSD ansatz for different molecules. Here O describes the number of orbitals (also the number of qubits) and P represents the number of parameters that have to be used by the q-UCCSD ansatz to cover the Hilbert space for the calculation of ground state energy.}
  
  \resizebox{\textwidth}{!}{
  \begin{tabular}{|c|c|c|c|c|c|c|c|c|c|c|c|c|}
    \hline
   \multirow{2}{*}{Basis}  & \multicolumn{2}{c|}{$H^{-}$} & \multicolumn{2}{c|}{$OH^{-}$}  &  \multicolumn{2}{c|}{\ce{H2O}} & \multicolumn{2}{c|}{\ce{H2}}  & \multicolumn{2}{c|}{HF} & \multicolumn{2}{c|}{$F^{-}$}\\
    \cline{2-13}
   \multirow{2}{*}{} &O&P&O&P&O&P&O&P&O&P&O&P\\
    
    
    \hline   
      3-21G & 4 & 3& 22 & 1260 & 26 & 2240 & 8& 15 & 22 & 1260 & 18& 560\\
    \hline  
    6-31G & 4 & 3& 22 & 1260 & 26 & 2240 & 8& 15 & 22 & 1260 & 18& 560 \\ 
    \hline
      6-311+$G^{*}$ & 6 & 8 & 50 & 14000 & 56 & 18515 & 12 & 35 & 50 & 14000 & 44 & 10115\\
     \hline
     6-31++G & 
  6 & 8 & 32  & 4235 & 38 &  6860 & 12 & 35 & 32 & 4235 & 26 & 2240\\
     \hline 
     6-31++$G^{**}$ & 12 & 35 & 48 & 12635 & 60  & 21875 & 24 & 143 & 48 & 12635 & 36 & 5915\\
    \hline
    aug-cc-pVDZ & 18 & 80 & 64 & 25515  &  82 & 45360 & 36 & 323 & 64 & 25515 & 46 & 11340\\
    \hline
  \end{tabular}
  }
  \label{tbl: number of orbital and parameters}
\end{table}

 For \ce{H2O} we have simulated the symmetrical stretching of the O-H bond with a fixed angle of $104.51^{0}$. Table \ref{tbl: all molecules 1} represents the ground state energy values for the molecules with different basis and methods. Although all the molecules we have taken are singlet we were unable to simulate all of them in minimal basis sets (STO-3G, STO-6G). For example, $F^{-}$ has 10 electrons and when we take a minimal basis set, the orbital used by the molecule is exactly 10. A similar problem is with the hydride ion also, it is a two-electron system and uses two orbitals in the minimal basis sets which do not allow it to take part in any single or double excitation. Therefore, we use the split valence basis functions\cite{ditchfield1971self} (3-21G, 6-31G, and 631++G) for our calculation. Furthermore, with higher basis sets the number of qubits and parameters increases which does not allow us to simulate the bigger molecules such as \ce{H2O},$OH^{-}$ and HF (Table \ref{tbl: number of orbital and parameters}). Water molecule will need about 82 qubits, hydroxide will need 64 qubits and hydrogen fluoride will need about 64 qubits if we try to simulate in the basis set where the hydride ion is giving the best value (Table \ref{tbl: number of orbital and parameters}). From Table \ref{tbl: all molecules 1} we can see that the q-UCCSD ansatz shows good agreement with respect to the FCI and CCSD energy but the pair q-UCCD ansatz is unable to converge to the ground state energy as compared to FCI. A similar case is there for the RY-CX ansatz, although it uses fewer parameters and CNOT gate and is the most consistent ansatz for our hydride ion simulation, it is also unable to get the accuracy with respect to the FCI energy. This concludes that the hardware-efficient ansatz does not depend on the type of molecular systems for calculation and is not fully reliable for these calculations.

\begin{table}[H]
 \resizebox{\textwidth}{!}{
    \centering
    \begin{tabular}{|c|c|c|c|c|c|c|c|}
    \hline
   Molecule & Basis sets  & Hartree-Fock  & CCSD  &  FCI  & RY-CX & pair q-UCCD & q-UCCSD \\
    & & (HF) Energy & Energy & Energy& (Full)& &\\
    \hline
       
    \multirow{3}{*}{ $OH^{-}$ }  & 3-21G & -74.86721695 & -74.98390701 & -74.98525103  & -72.99530259 & -74.91093858 & -74.87810391 \\
   \cline{2-8}
        & 6-31G & -75.31175321 & -75.44266001 & -75.44411134 & -73.33991053 & -74.99093858
        & -75.44006001
        \\
   \cline{2-8}
        & 6-31++G & -75.36290608 & -75.51786362 & -75.52224202 & -73.62598631 & -75.10001254 & -75.51002694\\
        
    \hline
\multirow{3}{*}{ $H_{2}O$ }  & 3-21G & -75.58488347 & -75.71719507 & -75.71896929  & -70.82694187 & -75.60023677 & -75.58595909 \\
   \cline{2-8}
        & 6-31G & -75.98165793 & -76.12012519 & -76.12179749 & -70.99694187 & -75.99974658 & -76.11882801 \\
   \cline{2-8}
        & 6-31++G & -76.10025621 & -76.24087269 & -76.25100255 & -71.45259751 & -75.99984232& -76.13226809\\
     \hline   
    
    \multirow{3}{*}{ $H_{2}$}  & 3-21G & -1.12296002 & -1.14787700 & -1.14770817  & -1.14073954 & -1.13977224 & -1.14770797 \\
   \cline{2-8}
        & 6-31G & -1.12682800 & -1.15162500 & -1.15152108  & -1.33436724 & -1.14342063 & -1.15161427 \\
   \cline{2-8}
        & 6-31++G & -1.12692377 & -1.151731061  & -1.15173111 & -0.97388904 & -1.14075886 & -1.15173105\\
     \hline   

\multirow{3}{*}{ HF }  & 3-21G & -99.45991942 & -99.58599051 & -99.58680834  & -98.67393645 & -99.47438766 & -99.47248562 \\
   \cline{2-8}
        & 6-31G & -99.98342600 & -100.11489030 & -100.1159399  & -98.99238754 & -99.58301937 & -99.98800342 \\
   \cline{2-8}
        & 6-31++G & -99.98992600 & -100.13492900& -100.1367459 & -99.10236999 & -99.68500012&  -99.99565231\\
     \hline   
   
\multirow{3}{*}{ $F^{-}$ }  & 3-21G & -98.77213500 & -98.88209161 & -98.88235445  & -94.19719723 & -98.78269256 & -98.79807546 \\
   \cline{2-8}
        & 6-31G & -99.35018060 & -99.47376951 & -99.47434888 & -94.99245893 & -99.36384366 & -99.38015238\\
   \cline{2-8}
        & 6-31++G & -99.41737596 & -99.56645761& -99.57012909 & -95.77149263 & -95.98535984& -99.41737596 \\
     \hline   

    \multirow{3}{*}{ $H^{-}$ }  & 3-21G & -0.40042019 & -0.40850110 & -0.40850110 & -0.49619848 & -0.40843457 & -0.40850109\\
   \cline{2-8}
       & 6-31G & -0.42244193 & -0.43137771& -0.43137772 & -0.49823275& -0.43129428 & -0.43137771\\
  \cline{2-8}
        & 6-31++G & -0.48707258 & -0.51001427& -0.51001427& -0.49980979& -0.50243001& -0.51001421 \\
     \hline      
    \end{tabular}
    }
    \caption{The values of ground state energy across the basis sets for molecules participating in the proton transfer reaction. HF represents the hydrogen fluoride molecule. From classical computational ansatz, the Hartree-Fock, CCSD, and FCI methods are used. From the quantum computational ansatz, the values from two chemistry-inspired ansatz i.e. q-UCCSD, pair q-UCCD, and one HEA (RY-CX with full entanglement) are obtained. All the values are calculated in the units of Hartree (Ha) }
    \label{tbl: all molecules 1}
\end{table}

The reaction energies are calculated by subtracting the energy value of the reactant from the product as:

\begin{equation}
    \label{eq:reaction energy}
    \Delta E_{r}= Energy(product)-Energy(reactants)           
\end{equation}

For better comparison, the experimental energies of each of the molecules are taken the same as taken by Calvin et.al.\cite{ritchie1968theoretical1,ritchie1968theoretical2} The experimental energy of molecules, \ce{H2O}, $H^{-}$,$OH^{-}$ , $H_{2}$, HF and $F^{-}$ are -76.438 Hartree \cite{ritchie1967gaussian}, -0.526 Hartree\cite{roothaan1960correlated}, -75.804 Hartree \cite{ritchie1967gaussian}, -1.174 Hartree\cite{roothaan1960correlated}, -100.45 Hartree and -99.86 Hartree \cite{scherr1962perturbation} respectively.

\begin{table}[h]
  \caption{The reaction energy proton transfer reaction 1 ($H^{-}+H_{2}O\rightarrow OH^{-}+H_{2}$) and proton transfer reaction 2 ($H^{-}+HF\rightarrow F^{-}+H_{2}$). For each method, we have calculated the energies of all molecules involved in that reaction in each basis set. To calculate the ground state energy for the first reaction, we used the formula $\Delta E_{r}$= $(E_{H_{2}}+E_{OH^{-}})$-$(E_{H^{-}}+ E_{\ce{H2O}})$
  and for the second reaction we used the formula
  $\Delta E_{r}$= $(E_{H_{2}}+E_{F^{-}})$-$(E_{H^{-}}+ E_{HF})$ . All the values of reaction energies are in the units of Hartree (Ha).}
  \label{tbl:hydride ion with water molecule}
  \resizebox{\textwidth}{!}{
  \begin{tabular}{|c|c|c|c|c|c|c|}
  \hline
   \multirow{2}{*}{Method} & \multicolumn{3}{c|}{$H^{-}+H_{2}O\rightarrow OH^{-}+H_{2}$} & \multicolumn{3}{c|}{$H^{-}+HF\rightarrow F^{-}+H_{2}$}\\
    \cline{2-7}
      & $\Delta E_{r}$ & $\Delta E_{r}$ & $\Delta E_{r} $ & $\Delta E_{r}$ & $\Delta E_{r}$ & $\Delta E_{r} $\\
    
   \multirow{2}{*}{} & (3-21G) & (6-31G) & (6-31++G)& (3-21G) & (6-31G) & (6-31++G) \\
    
    \hline
    
     Hartree-Fock & -0.0048732819 & -0.0344813496 & 0.0974989493 & -0.0347553819 & -0.0711406696 & -0.067301141 \\
     \hline
    
     CCSD & -0.0060878389 & -0.0427821035 & 0.0812922877 & -0.0347531829 & -0.0785523426 & -0.0732453923 \\
    
     \hline
      FCI & 0.0054888129 & -0.0424572126 & -0.0129563013 & -0.0347531829 & -0.0785523426 &  -0.0751000223  \\
      \hline
      
      RX-RY-Full & -2.812901786 &  -3.179103152 & -2.647468054 & 3.832198155 &  3.163794118 &  2.85679811 \\
      \hline 
      pair q-UCCD & -0.04203947951 & 0.2966816405 & 0.2615009277 & -0.0396425716 &  -0.4929506479 & 3.061311428   \\
     \hline
      q-UCCSD & -0.03135169009 & -0.0414685544 &   -0.0194614393 &  -0.0647967101 & -0.1123855144 & -0.0634404853 \\
     \hline
  \end{tabular}
  }
\end{table}

By taking the experimental energies for each molecule and using the Eq. \ref{eq:reaction energy}, the hydride ion reacting with water to form hydrogen and hydroxide reaction is exothermic by 8.8 kcal(-0.014 Ha), and the hydride ion reacting with hydrogen fluoride to form hydrogen and fluoride ion is exothermic by -0.058 Ha.  Calvin et .al. have calculated the reaction energy for hydride ion reacting with water to be -0.054 Ha and 0.006 Ha with respect to the small and large basis. Similarly, for the hydride ion reacting with hydrogen fluoride, the reaction energy was calculated as 0.018 Ha and -0.057 Ha with respect to the small and large basis set\cite{ritchie1968theoretical1,ritchie1968theoretical2}.
The main component here for all our calculations is the hydride ion. Because for the reaction energy all the molecules should show accurate values to their experimental energy but the hydride ion is unable to show that behavior at lower basis sets. That is the reason the reaction energy for the basis 3-21G and 6-31G is not up to the mark with any of the methods as the value of hydride ion with this method is not good. However, when we are taking basis set 6-31++G the hydride ion ground state improves and led the reaction energy to a better point.
with this basis, the classical computational methods FCI and CCSD methods the energies obtained are -0.0129563013 Ha (exothermic) and 0.081292287 Ha (endothermic) for the first proton transfer reaction ($H^{-}+H_{2}O\rightarrow OH^{-}+H_{2}$). Similarly, for proton transfer reaction -2 ($H^{-}+HF\rightarrow F^{-}+H_{2}$) the values of reaction energy with FCI and CCSD method are -0.07510023 Ha (exothermic) and -0.0732453923 Ha (exothermic). 
With the q-UCCSD ansatz the best value of the reaction energy for proton transfer reaction-1(reaction of hydride ion with water) we get the value of -0.019461439 Ha which is close to the energy obtained through the experimental value of each molecule. Similarly, the value for the proton transfer reaction-2  (reaction of hydride ion with hydrogen fluoride) with q-UCCSD ansatz is ~ -0.0634404 Ha showing good agreement with the experimental. Whereas for the other two ansatzes, RY-CZ full entanglement and pair q-UCCD the reaction energies in both cases are endothermic in nature and are in bad agreement with the energy obtained through the experimental value of each molecule.

\section{Conclusion}

In summary, we investigated the effectiveness and accuracy of the quantum computational ansatz (q-UCCSD) by doing a comparison with the classical computational ansatz. We studied a highly correlated hydride ion that has a wide range of applications in astrophysics and quantum chemistry.
Utilizing a range of quantum computational methods (chemistry-inspired and HEA)  within the VQE formulation, we aimed to recover more electron correlation energy. To achieve this goal, we systematically employed increasingly larger and more flexible basis sets.
As a result, the q-UCCSD ansatz calculated the ground state energy and single-electron detachment energy of the hydride ion with much higher precision. 
The q-UCCSD ansatz showed very accurate results within the chemical accuracy,  with a percentage relative error of $10 ^{-5}$ with respect to FCI and CCSD energy. The calculated ground state energy  matches closely with the calculation done by Chandrasekhar. Additionally, we  studied the behavior of two other ansatzes based on the unitary coupled cluster i.e. singlet q-UCCD and pair q-UCCD, which failed to depict the accurate correlation energy of the hydride ion. The HEA ansatzes  (RX-RY-CZ with
linear entanglement, RX-RY-CZ with full entanglement, and RY-CX with full entanglement), are easy to implement for the calculation. However, as the number of orbitals increases
it fails to converge to the ground state energy. Using extra quantum resources for extending the basis set produced results that are more precise but required a large quantum computational cost. We got the results for single-electron detachment energy of hydride ion using the q-UCCSD ansatz (0.024694149 Ha), which closely matches the experimental value obtained by the laser-ion beam photodetachment apparatus. Additionally, the calculation of reaction energy for two proton transfer reactions allowed us to check the behavior of both chemistry-inspired and hardware-efficient ansatzes when a number of molecules take part in a reaction simultaneously. Our results showed that the reaction energy of the hydride ion with water forming hydroxide, and the reaction energy of the hydride ion with hydrogen fluoride to produce fluoride ion are exothermic in nature which is an improved result from its previous work. In conclusion, the combination of the q-UCCSD ansatz and VQE have been shown to be a powerful and precise method for computing unstable hydride ion. It effectively mimics the classical coupled cluster, which allows it to recover the required correlation energy.


\begin{acknowledgement}
We would like to thank IISER Kolkata for providing hospitality during the course of the project work and
QUEST(DST/ICPS/QuST/Theme-1/2019/2020-21/01) for financial support. We thank Mr. Rajiuddin, Mr. Dheerendra, Miss Sangeeta, from IISER Kolkata, and Miss Suprava from the Central University of Punjab for numerous enlightening discussions throughout the project.

\end{acknowledgement}



\bibliography{ref.bib}

\end{document}